\documentclass{jaa}
\usepackage{natbib}
\usepackage{graphicx}
\usepackage{newtxtext,newtxmath}
\usepackage{url}
\usepackage[T1]{fontenc}
\usepackage{amsmath}	
\usepackage{hyperref}
\usepackage{xspace}
\usepackage{fixltx2e}

\begin{document}\sloppy

\title{X-ray polarization observations of IC 4329A with IXPE: Constraining the geometry of the 
X-ray corona}


\author{Indrani Pal\textsuperscript{1,*}, C. S. Stalin\textsuperscript{1}, Rwitika Chatterjee\textsuperscript{2}, Vivek 
K. Agrawal\textsuperscript{2}}
\affilOne{\textsuperscript{1}Indian Institute of Astrophysics, Block II, Koramangala, Bangalore 560 034, India\\}
\affilTwo{\textsuperscript{2}Space Astronomy Group, ISITE Campus, U. R. Rao Satellite Centre,  Bangalore 560 037, India.}


\twocolumn[{

\maketitle

\corres{indrani.pal@iiap.res.in}

\msinfo{1 January 2015}{1 January 2015}

\begin{abstract}
X-ray polarimetry is a powerful tool to probe the geometry of the hot X-ray corona in active galactic nuclei (AGN). Here, we present our results on the characterisation of the X-ray polarization of the radio-quiet  Seyfert-type AGN IC 4329A at a redshift of $z$ = 0.016. This is based on observations carried out by the {\it Imaging X-ray Polarimeter (IXPE)}. {\it IXPE} observed IC 4329A on January 5, 2023, for a total observing time of 458 ks. From the model-independent analysis, we found a polarization degree ($\Pi_{X}$) of 3.7$\pm$1.5$\%$ and a polarization position angle ($\Psi_{X}$) of 61$^{\circ}$$\pm$12$^{\circ}$ in the 2$-$8 keV energy range (at 1$\sigma$ confidence level). This is also in agreement with the values of $\Pi_{X}$ and $\Psi_{X}$ of 4.7$\pm$2.2$\%$ and 71$^{\circ}$ $\pm$14$^{\circ}$ respectively obtained from spectro-polarimetric analysis of the I, Q and U Stokes spectra in the 2$-$8 keV energy band (at the 90$\%$ confidence). The value of $\Pi_X$ in the 2-8 keV band obtained from the model-independent analysis is lower than the minimum detectable polarization (MDP) value of 4.5$\%$. However, $\Pi_X$ obtained from spectro-polarimetric analysis in the 2-8 keV band is larger than the MDP value. In the 3-5 keV band, we found $\Pi_X$ of 6.5 $\pm$ 1.8, which is larger than the MDP value of 5.5$\%$. The observed moderate value of $\Pi_{X}$ obtained from the analysis of the {\it IXPE} data in the 3$-$5 keV band argues against a spherical lamp$-$post geometry for the X-ray corona in IC 4329A; however, considering simulations, the observed polarization measurements tend to favour a conical shape geometry for the corona. This is the first time measurement of X-ray polarization in IC 4329A. Measurements of the X-ray polarization in many such radio-quiet AGN will help in constraining the geometry of the X-ray corona in AGN.
\end{abstract}


}]


\doinum{12.3456/s78910-011-012-3}
\artcitid{\#\#\#\#}
\volnum{000}
\year{0000}
\pgrange{1--}
\setcounter{page}{1}
\lp{1}

\section{Introduction}
Active galactic nuclei (AGN), one among the luminous objects in the Universe, are believed to be powered by the accretion of matter onto super-massive black holes (SMBHs;$\sim$ 10$^6$ $-$ 10$^{10}$ $M_{\odot}$) situated at the centres of galaxies \citep{1969Natur.223..690L,1973A&A....24..337S, doi:10.1146/annurev.aa.31.090193.002353, 1984ARA&A..22..471R, 1995PASP..107..803U}. They emit over the entire accessible electromagnetic spectrum, such as the high energy gamma-rays \citep{1999ApJS..123...79H}, X-rays \citep{2012ARA&A..50..455F}, UV \citep{1995ApJ...451..498M}, optical \citep{1992ApJS...80..109B}, infrared \citep{1996ARA&A..34..749S} and radio \citep{1991ApJ...378...65C}. The AGN come under two broad categories, a majority of them emit less or no radio emission and are called radio-quiet AGN, while a minority of about 10 per cent emit copiously in the radio band and are called radio-loud AGN \citep{1989AJ.....98.1195K}. Though X-rays are emitted by the different classes of AGN, the detailed physics of the cause of X-ray emission is less understood and is highly debated. Also, the relative contribution of the physical processes to the X-ray emission may be different between the different classes of AGN.

\begin{figure}
        \includegraphics[scale=0.45]{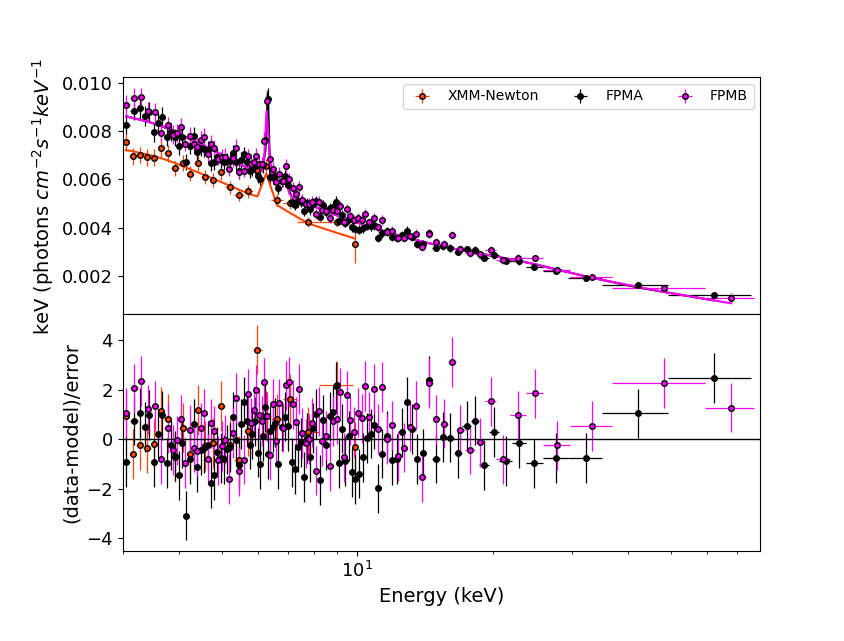}
    \caption{The {\it XMM-Newton EPIC PN}, {\it NuSTAR} FPMA and FPMB spectra fitted jointly in 3$-$78 keV energy band using the model {\it const$\times$TBabs$\times$zTBabs$\times$(xillverCP)}. The observations were taken simultaneously on August 12, 2021.}
    \label{figure-7}
\end{figure}

In the radio-quiet category of AGN, the observed X-ray emission is believed to be due to inverse Compton scattering of optical-UV radiation from the accretion disk by hot electrons ($\sim10^{8-9}$ K) in the corona, a constituent of AGN \citep{1991ApJ...380L..51H, 1993ApJ...413..507H, 1994ApJ...432L..95H}. The X-ray corona is believed to be located close to the accretion disk, with microlensing studies pointing it to be compact with the size of a few gravitational radii ($r_g = GM/c^2$;\citealt{2016AN....337..356C}). X-ray reverberation mapping studies also indicate the corona to be compact, of the order of 3$-$10 $r_g$ in radius \citep{2009Natur.459..540F}. 

Recent studies, most notably using observations from the {\it Nuclear Spectroscopic Telescope Array} ({\it NuSTAR}, \citealt{2013ApJ...770..103H}) were able to characterize the nature of the corona in AGN, such as its temperature ($\rm{kT_e}$) and variation in its temperature \citep{2017ApJ...836....2Z, 2016MNRAS.463..382U, 2015A&A...577A..38U, 2016MNRAS.456.2722K, 2021MNRAS.502...80K, 2022A&A...662A..78P, 2022MNRAS.517.3341P, 2023MNRAS.518.2529P}. This is achieved via modelling of their broadband X-ray spectra, as the shape of the primary X-ray continuum in AGN is sensitive to the temperature of the coronal plasma, its optical depth, the geometry of the corona and the nature of the seed photons \citep{1993ApJ...413..507H, 1994ApJ...432L..95H}.

The corona could be either a spherical structure above the black hole (spherical lamp-post; \citealt{2017MNRAS.467.2566F}) or a slab-like structure sandwiching the accretion disk \citep{1993ApJ...413..507H} or a conical structure \citep{10.1093/mnras/stab3745}. From the physical model fits to the observed high energy spectra of AGN, the $\rm{kT_e}$  reported in the literature assumes the corona to have either a spherical or a slab geometry \citep{2018ApJ...856..120R}. Also, from the multi-epoch spectral analysis available in the literature, several models of the corona, including changes in the structure of the corona, are proposed to explain the observed changes in $\rm{kT_e}$. \citep{2022A&A...662A..78P,2023MNRAS.518.2529P}. However, from spectral analysis of AGN, it is difficult to distinguish between different coronal geometries \citep{2018A&A...614A..37T,2019A&A...630A.131M}. Irrespective of several studies in the literature attempting to characterise the corona in AGN, we still lack knowledge on (i) the origin of the corona, (ii) the cause of its high temperature and (iii) its geometry.

X-ray polarimetric observations on AGN can yield the needed measurement to constrain the geometry of the X-ray-emitting corona in AGN \citep{1989ESASP.296..991M,1993MNRAS.264..839M}. This is because polarization depends on the geometry of the emitting region and of the external photon field \citep{2018A&A...619A.105T,Zhang_2019}. The launch of  the  {\it Imaging X-ray Polarimetry Explorer} ({\it IXPE}; \citealt{10.1117/1.JATIS.8.2.026002}) on December 9, 2021, sensitive in the 2$-$8 keV band has enabled the studies of X-ray polarization from AGN. As of now, {\it IXPE} has observed four radio-quiet AGN, namely MCG-05-23-16, the Circinus galaxy, NGC 4151 and IC 4329A. Of these, for MCG-05-23-16, \cite{2022MNRAS.516.5907M}, reported a polarization degree $\Pi_{X}$ $<$ 4.7$\%$. In the case of the Circinus Galaxy, \cite{2023MNRAS.519...50U}, reported a high value of $\Pi_{X}$ = 28$\%$, which is thought to be due to reflection from the torus. For NGC 4151, \cite{2023arXiv230312541G}, found values of $\Pi_{X}$ and $\Psi_{X}$ of 4.9 $\pm$1.1$\%$ and 86$^{\circ}$$\pm$7$^{\circ}$ respectively, arguing for a slab geometry of the corona. Here, we report the results of polarization on the fourth radio-quiet AGN, IC 4329A.

IC 4329A with a black hole mass of 6.8$^{+1.2}_{-1.1}$ $\times$ 10$^7$ M$_{\odot}$ \citep{Bentz_2023} is one of the brightest  Seyfert 1.2 galaxies \citep{2010A&A...518A..10V} in X-rays at a redshift z = 0.016 \citep{1991AJ....101...57W} with an X-ray spectrum having many absorbing systems \citep{2005A&A...432..453S}. It is an ideal object to study the intrinsic continuum and reflection from any material surrounding the source. It has been studied extensively for its X-ray spectral and timing properties \citep{10.1093/mnras/stab1113}.  Observations of IC 4329A in the millimetre were explained as due to synchrotron emission in the hot corona \citep{2018ApJ...869..114I}. From Comptonization model fits to the {\it NuSTAR} observations on IC 4329A, \cite{2016MNRAS.458.2454L}, report a $\rm{kT_e}$ of $40^{+7}_{-5}$ keV using the {\tt COMPPS} model, assuming a spherical geometry of the corona. Alternatively, \cite{2001ApJ...556..716P}, using the Comptonization model and a slab geometry of the corona, found a value of $\rm{kT_e}$ = $170^{+10}_{-5}$ keV. \cite{2018A&A...614A..37T} reported $\rm{kT_{e}}$ = 37$\pm$7 keV from fitting {\it compTT} to the source spectrum for slab geometry. \cite{Kang_2022} estimated $\rm{kT_{e}}$ = 71$^{+37}_{-15}$ keV using {\it relxillCP} model. \cite{2022ApJ...927...42K} also found $\rm{kT_{e}}$ = 82$^{+16}_{-7}$ keV from {\it xillverCP} fit to the source spectrum. We analyzed the {\it NuSTAR} spectrum taken in 2012. Fitting the spectrum using {\it xillverCP}, we obtained  best-fit values of $\Gamma$ and $\rm{kT_{e}}$ of 1.83$^{+0.003}_{-0.003}$ and 64$^{+15}_{-12}$ keV respectively (Pal et al. 2023, under preparation). From a joint fit of {\it XMM-Newton EPIC PN} and {\it NuSTAR} spectra observed simultaneously on August 12, 2021, we found $\Gamma$ = 1.88$^{+0.01}_{-0.01}$ and $\rm{kT_{e}}$ $>$ 135 keV. The best-fit model with data and the residues are given in Fig. \ref{figure-7}. All the spectral fits produced good fit statistics with $\chi^{2}/dof$ $\sim$ 1.0, irrespective of the coronal geometry assumed in the used models. Thus, from analysis of the AGN X-ray continuum, it is not possible to constrain the geometry of the corona, while X-ray polarimetric observations could provide the needed constraint on the coronal geometry.  Here, we present our results on the analysis of the first X-ray polarimetric observations carried out on this source by {\it IXPE}. The observations and data reduction are described in Section 2, the analysis is described in Section 3, and the results are discussed in Section 4, followed by the summary in the final section.

\begin{figure}
        \includegraphics[scale=0.63]{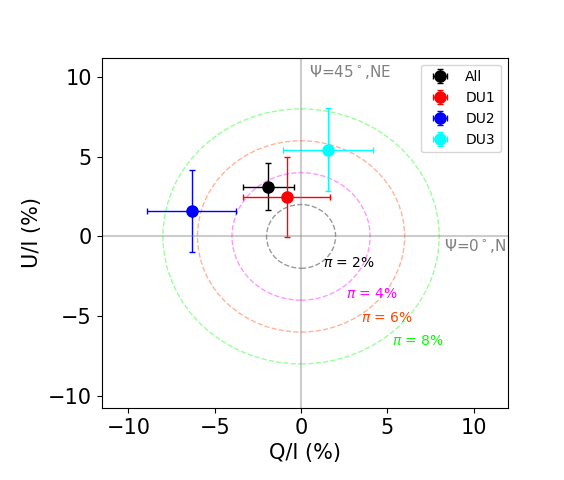}
    \caption{Normalized U/I and Q/I Stokes parameter in the total 2$-$8 keV band of
{\it IXPE}. The plotted errors are at the 1$\sigma$ uncertainties, and the concentric
circles correspond to different values of polarization degree.}
    \label{figure-1}
\end{figure}

\section{Observations and data reduction}
{\it IXPE} \citep{10.1117/1.JATIS.8.2.026002} observed IC 4329A on January 5, 2023, with its three detector units (DUs) for a net exposure of about 458 ks. The log of the {\it IXPE} observation is given in Table \ref{table-1}. The calibrated data were produced by the standard {\it IXPE} pipeline provided by the Science Operation Center (SOC)\footnote{\href{https://heasarc.gsfc.nasa.gov/docs/ixpe/analysis/IXPE-SOC-DOC-009-UserGuide-Software.pdf}{https://heasarc.gsfc.nasa.gov/docs/ixpe/analysis/IXPE-SOC-DOC-009-UserGuide-Software.pdf}}. We used the cleaned and calibrated level 2 data for the scientific analysis.

\begin{figure*}
\hbox{
\hspace*{-4.0cm} \includegraphics[scale=0.33]{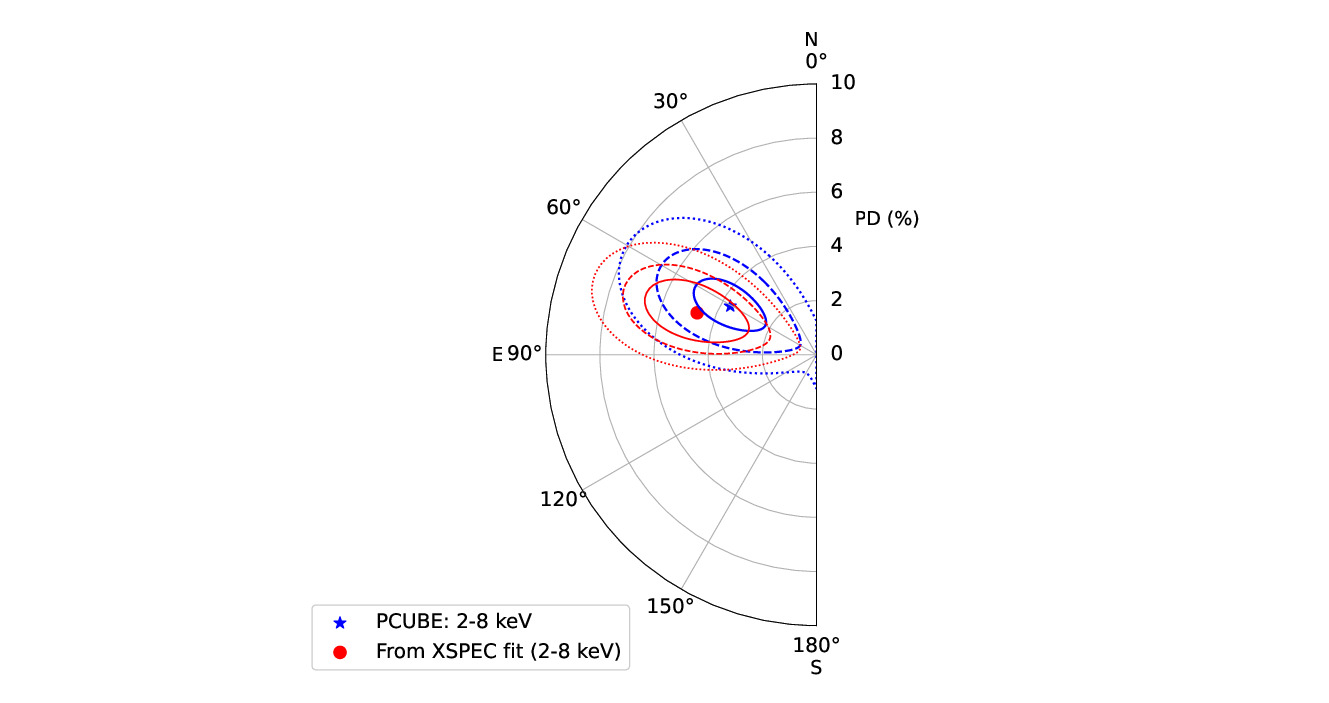}
\hspace*{-4.0cm} \includegraphics[scale=0.33]{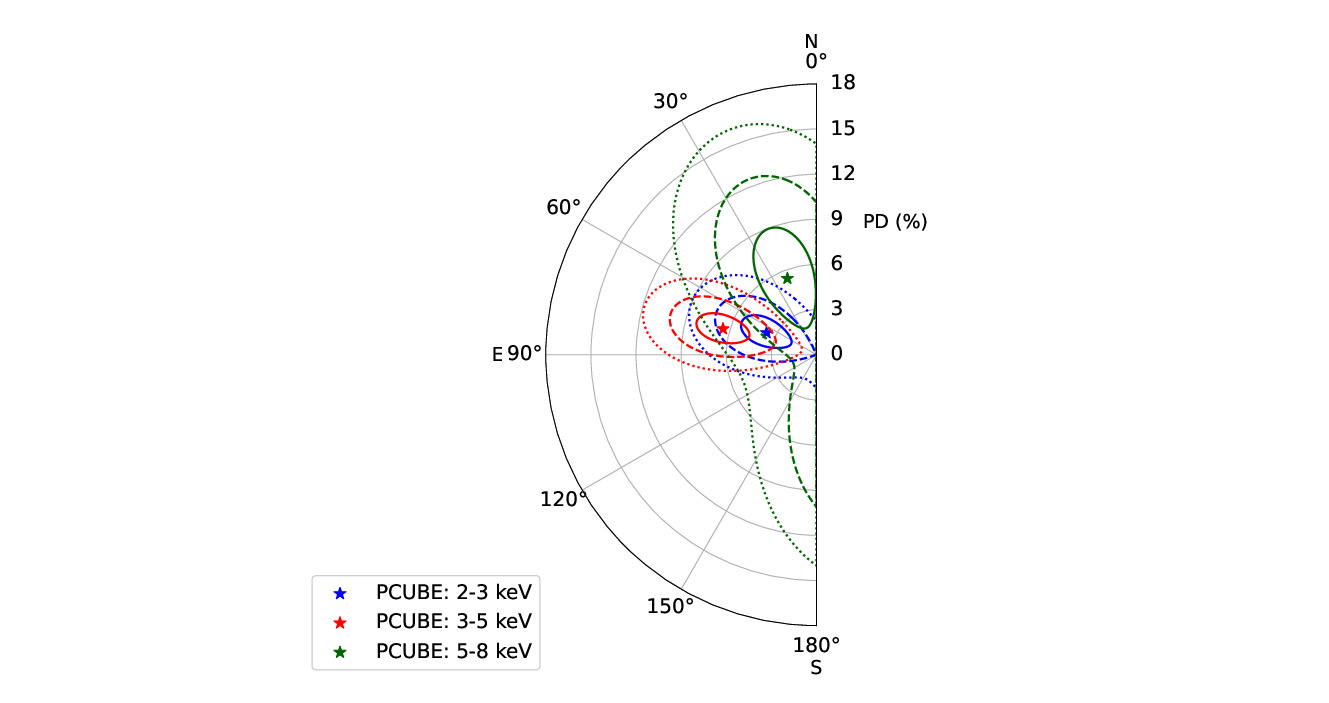}
     }
\caption{The 68$\%$, 90$\%$ and 99$\%$ confidence contours between $\Pi_{X}$ and $\Psi_{X}$ in the 2$-$8 keV (left panel) band and in three different energy bands (right panel).}
\label{figure-2}
\end{figure*}

The publicly available level 2 data were analysed using {\tt IXPEOBSSIM} software v30.0.0 \citep{BALDINI2022101194}. A count map in sky coordinates was generated using the {\tt CMAP} algorithm within the {\tt xpbin} task. We adopted a circular region with a radius of $70''$ for the source extraction from the three DUs, and a source-free region with a radius of $100''$ was chosen for the background extraction for each DU. We then used the {\tt xpselect} task to generate the filtered source and background regions. For the spectro-polarimetric analysis, the I, Q and U source and background spectra were generated using the {\tt PHA1}, {\tt PHA1Q} and {\tt PHA1U} algorithm using {\tt xpbin} task within {\tt IXPEOBSSIM} for the three DUs. We used a minimum of 30 counts/bin to bin the I spectra, where a constant energy binning of 0.2 keV was used for Q and U spectra.

\begin{table}
        \centering
        \caption{Log of {\it IXPE} observation.}
        \label{table-1}
\begin{tabular}{ccc}
\hline
OBSID  & Date & Exposure Time \\
      &       &        (secs)         \\
\hline
01003601 & 2023-01-05 & 457715  \\
        \hline
        \end{tabular}
\end{table}

\section {Analysis}
\subsection{Polarimetry}
The polarimetric signal from IC 4329A was analysed using the {\tt PCUBE} algorithm in the {\tt xpbin} task. The three polarization cubes for the three DUs were generated to extract information like the Stokes parameters (I, Q, U); the minimum detectable polarization (MDP); the polarization degree ($\Pi_{X}$); the polarization angle ($\Psi_{X}$) and their associated errors. We first generated the three polarization cubes corresponding to three DUs in the entire 2$-$8 keV energy band. The combined polarization parameters from the three DUs in the 2$-$8 keV band are given in Table \ref{table-2}. We found $\Pi_{X}$ = 3.7$\pm$1.5$\%$, MDP = 4.5$\%$ and $\Psi_{X}$ = 61$^{\circ}$$\pm$12$^{\circ}$. The normalized U/I and Q/I Stokes parameters obtained from the three polarization cubes corresponding to the three DUs and from the combined cube are shown in Fig. \ref{figure-1}.

To check for the energy dependence of the polarization parameters, we also derived the polarization parameters in three energy bins of 2$-$3 keV, 3$-$5 keV and 5$-$8 keV using the {\tt PCUBE} algorithm. The derived parameters are given in Table \ref{table-2}. The polarization contours for the total energy range of 2$-$8 kev as well as for the different energy ranges, are given in Fig. \ref{figure-2}. From Table \ref{table-2} and Fig. \ref{figure-2}, it is evident that $\Pi_{X}$ and $\Psi_{X}$  are consistent within errors in the energy range of 2$-$3, 3$-5$ and the 2$-$8 keV, while there is a tendency for a decrease in $\Psi_{X}$ in the 5$-$8 keV band. However, the error is so large to arrive at any firm conclusion on the change in $\Psi_{X}$ at the higher energy band. 

Though the values of $\Pi_X$ in the 2$-$3, 5$-$8 and 2$-$8 keV bands are close to or lower than the MDP values, at 3$-$5 keV, we found a $\Pi_X$ of 6.5$\pm$1.8$\%$ which is larger than the MDP value of 5.5$\%$. We thus conclude to have detected significant X-ray polarization in IC 4329A, and this is the first report of the detection of the X-ray polarization signal in IC 4329A.

\subsection{Spectro-polarimetry}
We carried out the spectro-polarimetric analysis of the {\it IXPE} I, Q, U spectra in the 2$-$8 keV energy band.
For the spectral fitting, we used an absorbed {\it powerlaw}, modified by a multiplicative constant polarization model {\it polconstant} in XSPEC V12.13.0c \citep{1996ASPC..101...17A}. This model assumes a constant polarization degree and a constant polarization angle over the specific energy band. In XPSEC the model takes the following form,
\begin{equation}
    constant \times TBabs \times zTBabs \times (polconst \times po)
\end{equation}

\begin{figure*}
\hbox{
     \includegraphics[scale=0.55]{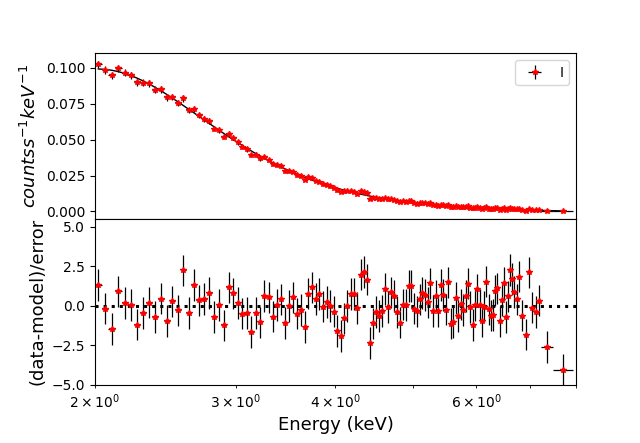}
     \includegraphics[scale=0.55]{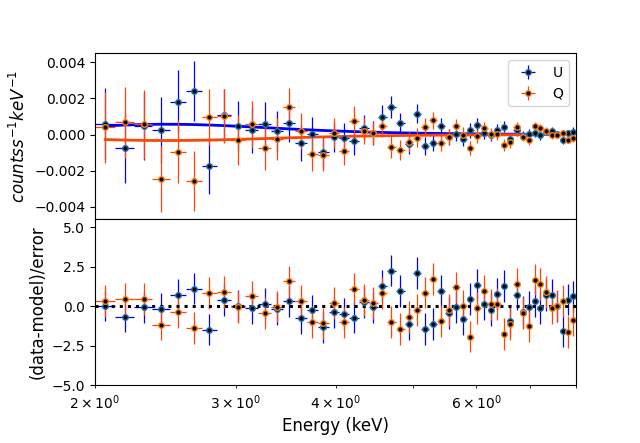}
     }
\caption{Left panel: {\it IXPE} I Stokes best-fit spectra with residuals. Right panel: Q and U stokes best-fit spectra with residuals.}
\label{figure-3}
\end{figure*}

Here, the {\it const} represents the inter-calibration constant for each detector, which varies between 0.95 to 1.00. {\it TBabs} was used to model the Milky Way Galactic hydrogen column density, which was taken from \cite{2013MNRAS.431..394W}. To model the host galaxy column density, we used {\it zTBabs} and let the column density ($N_{H}$) vary during the fit. This model fits the I, Q, and U spectra (from the three detectors) well with a reduced chi-square ($\Delta \chi^{2}$) = 1083/1064. The best fit I, Q and U spectra with the residues are given in Fig. \ref{figure-3}.  The spectro-polarimetric fit produced a $\Pi_{X}$ of 4.7$\pm$2.2$\%$ (larger than the MDP value of 4.5$\%$) and $\Psi_{X}$ of 71$^{\circ}$$\pm$14$^{\circ}$ associated with the primary emission when modelled with a power law of photon index ($\Gamma$) of 1.95$\pm$0.05. In Fig. \ref{figure-2}, the contours between $\Pi_{X}$ and $\Psi_{X}$ in 68$\%$, 90$\%$ and 99$\%$ are plotted along with the contours obtained from the polarimetric analysis. All the model parameters for each of the detectors were tied during the fit. The errors were calculated at the 90$\%$ confidence ($\chi^2$ = 2.71 criterion). The best-fit parameters are given in the Table \ref{table-3}. We also performed an MCMC analysis to calculate the errors in 90$\%$ confidence associated with the best-fit parameters. The errors from the MCMC analysis are also reported in Table \ref{table-3}.

\begin{table}
        \centering
        \caption{Polarization parameters in different energy bands.}
        \label{table-2}
\begin{tabular}{cccc}
\hline
Energy band & $\Pi_{X}\pm 1\sigma$ & MDP & $\Psi_{X}\pm 1\sigma$ \\
  keV    &  $\%$     &  $\%$    &  degree         \\
\hline
 2$-$ 8 & 3.7$\pm$1.5 & 4.5 & 61$\pm$12 \\
 2$-$ 3 & 3.6$\pm$1.8 & 5.5 & 66$\pm$14 \\
 3$-$ 5 & 6.5$\pm$1.8 & 5.5 & 74$\pm$8 \\
 5$-$ 8 & 5.4$\pm$3.5 & 10.73 & 21$\pm$19 \\
        \hline
        \end{tabular}
\end{table}

\section{Discussion}
The measured polarised X-ray emission in the radio-quiet category of AGN is believed to be due to the inverse Compton scattering of UV/optical accretion disk photons by
hot electrons in the corona. Therefore, the measured degree of X-ray polarization depends on the geometry of the corona. We examined the polarimetric properties of the Seyfert 1 galaxy IC 4329A using the {\it IXPE} data through a model-independent as well as spectro-polarimetric analysis. Using the model-independent approach in the 2$-$8 keV, we found a polarization angle of 61$^{\circ}$$\pm$12$^{\circ}$ and a polarization degree of 3.7$\pm$1.5$\%$ which is slightly lower than the MDP value of 4.5$\%$ (at the 1$\sigma$ confidence level). From spectro-polarimetric analysis, we found values of $\Pi_{X}$ and $\Psi_{X}$ of 4.7$\pm$2.2\% (larger than the MDP value of 4.5$\%$) and 71$^{\circ}$$\pm$14$^{\circ}$ respectively. Recently, \cite{10.1093/mnras/stab3745} simulated polarization signals expected from AGN corona for three different geometries, namely, slab (inner radius $\sim$ 10$R_{g}$), spherical lamp$-$post (radius $\sim$ 10$R_{g}$ and height $\sim$30$R_{g}$ above the accretion disk) and the truncated cone geometry for an outflowing corona (a failed jet with an outflowing velocity of 0.3c and initial radius of 20$-$30$R_{g}$). The authors claimed that for a slab coronal geometry, the polarization degree goes up to 12 per cent. For the symmetrical nature of a spherical lamp$-$post corona, it is expected that the polarization signal received from such sources would be lower than the slab one. For such geometrics, a very low polarization degree ($\sim$ 1$-$3 $\%$) is expected \citep{1996ApJ...470..249P, 2018A&A...619A.105T}. For the conical corona, the expected polarization could be between the slab and the spherical ones. From their simulation using the {\tt MONK} \citep{Zhang_2019} code,  \cite{10.1093/mnras/stab3745} also reported that the polarization angle is close to 180$^{\circ}$ for slab geometry, while for the spherical and conical geometries, it is scattered around 90$^{\circ}$. Our measured value of $\Pi_X$ = 6.5$\pm$1.8$\%$ in the 3-5 keV, larger than the MDP value of 5.5$\%$ rules out the spherical lamp$-$post geometry for the corona. However, compared with the simulations of \cite{10.1093/mnras/stab3745},  our measured values argue for a conical geometry for the corona in IC 4329A.

\section{Conclusions}
We carried out the analysis of the first X-ray polarimetric observations carried out by {\it IXPE}. The observation on IC 4329A for a total duration of 458 ksec was taken on 05 January 2023. The findings of the polarimetric study are as follows,
\begin{enumerate}
\item From a model-independent analysis, we found a value of $\Pi_{X}$ = 3.7$\pm$1.5\% and $\Psi_{X}$ = 61$^{\circ}$$\pm$12$^{\circ}$ in the 2$-$8 keV band. This value of $\Pi_X$ is slightly lower than the MDP value of 4.5$\%$.

\item From the spectro-polarimetric analysis of model fits to the spectra, we found values of $\Pi_{X}$ and $\Psi_{X}$ of 4.7$\pm$2.2\% and 71$^{\circ}$$\pm$14$^{\circ}$ respectively in the 2$-$8 keV band. Thus, the polarimetric measurements obtained from both model-independent and spectro-polarimetric analyses agree with each other. Also, the value of $\Pi_X$ obtained from the model fit to the spectrum is larger than the MDP value of 4.5$\%$.

\item To check for energy-dependent polarization, we derived the polarization parameters in different energy bands. While $\Pi_{X}$ is found to be similar within error bars in all the energy bands, the derived values of $\Psi_{X}$ are found to agree within errors in the 2$-$3, 3$-$5 and 2$-$8 keV bands. There is a tendency for lower values of $\Psi_{X}$ in the higher energy range of 5$-$8 keV. However, the error bar is too large in this energy range. Our observations do not find evidence of changes in polarization between energy bands. 

\item In the 3$-$5 keV band, from model-independent analysis, we found a $\Pi_X$ of 6.5$\pm$1.8$\%$ which is larger than the MDP value of 5.5$\%$. Also, from spectro-polarimetric analysis in the 2$-$8 keV band, we found a $\Pi_X$ of 4.7$\pm$2.2$\%$, which is larger than the MDP at that energy band. We, therefore, conclude to have detected X-ray polarization in IC 4329A.

\item Our observations, when compared with Monte-Carlo simulations in the literature \citep{10.1093/mnras/stab3745}, rule out a spherical lamp$-$post geometry for the corona in IC 4329A, instead tend to favour a corona with a conical geometry in IC 4329A.
\end{enumerate}

With the results reported in this work, the number of Seyfert type AGN with measured X-ray polarization measurements from {\it IXPE} observations have increased to four. X-ray polarimetric observations of more Seyfert-type AGN are needed to put better constrain the geometry of the X-ray corona in AGN. Also, repeated observations of the same source by {\it IXPE} would enable one to constrain for variations in the geometry of the corona, as has been hinted at from spectral modelling of the X-ray spectra of AGN.

\begin{table}
        \centering
        \caption{Results of spectro-polarimetric analysis.}
        \label{table-3}
\begin{tabular}{ccc}
\hline
Parameters & XSPEC analysis & MCMC analysis \\
\hline
$N_{H}$ & 0.70$^{+0.12}_{-0.12}$ & 0.70$^{+0.12}_{-0.11}$ \\
$\Pi_{X}$ ($\%$) & 4.7$^{+2.2}_{-2.2}$ & 4.7$^{+2.0}_{-2.4}$ \\
$\Psi_{X}$ ($^{\circ}$) & 71$^{+14}_{-14}$ & 71$^{+14}_{-15}$ \\
$\Gamma$ & 1.95$^{+0.05}_{-0.05}$ & 1.95$^{+0.05}_{-0.05}$ \\
$N$ & 0.023$^{+0.002}_{-0.002}$ & 0.023$^{+0.002}_{-0.002}$ \\
        \hline
        \end{tabular}
\end{table}

\section*{Acknowledgements}

The Imaging X-ray Polarimetry Explorer (IXPE) is a joint US and Italian mission. The US contribution is supported by the National Aeronautics and Space Administration (NASA) and led and managed by its Marshall Space Flight Center (MSFC), with industry partner Ball Aerospace (contract NNM15AA18C). The Italian contribution is supported by the Italian Space Agency (Agenzia Spaziale Italiana, ASI) through contract ASI-OHBI-2017-12-I.0, agreements ASI-INAF-2017-12-H0 and ASI-INFN-2017.13-H0, and its Space Science Data Center (SSDC) with agreements ASI- INAF-2022-14- HH.0 and ASI-INFN 2021-43-HH.0, and by the Istituto Nazionale di Astrofisica (INAF) and the Istituto Nazionale di Fisica Nucleare (INFN) in Italy. This research used data products provided by the IXPE Team (MSFC, SSDC, INAF, and INFN) and distributed with additional software tools by the High-Energy Astrophysics Science Archive Research Center (HEASARC), at NASA Goddard Space Flight Center (GSFC). Part of the French contribution is supported by the Scientific Research National Center (CNRS) and the French Space Agency (CNES). We thank the {\it NuSTAR} Operations, Software and Calibration teams for support with the execution and analysis of these observations. This research has made use of archival data of {\it XMM-Newton} and {\it NuSTAR} observatories through the High Energy Astrophysics Science Archive Research Center Online Service, provided by the NASA Goddard Space Flight Center. 

\vspace{-1em}






\bibliography{example}{}


\end{document}